\documentclass[lettersize,journal]{IEEEtran}
\usepackage{amsmath,amsfonts}
\usepackage{algorithmic}
\usepackage{algorithm}
\usepackage{array}
\usepackage[caption=false,font=normalsize,labelfont=sf,textfont=sf]{subfig}
\usepackage{textcomp}
\usepackage{stfloats}
\usepackage{url}
\usepackage{verbatim}
\usepackage{graphicx}
\usepackage{cite}
\begin{document}

\title{Deep Learning-Based CKM Construction with Image Super-Resolution}

\author{Shiyu Wang, Xiaoli Xu,~\IEEEmembership{Member,~IEEE,} Yong Zeng,~\IEEEmembership{Senior Member,~IEEE}
\thanks{Shiyu Wang and Xiaoli Xu are with the School of Information Science and Engineering, Southeast University, Nanjing 210096, China (e-mail: {220231205, xiaolixu}@seu.edu.cn). (Corresponding author: Xiaoli Xu.)}
\thanks{Yong Zeng is with the National Mobile Communications Research Laboratory, Southeast University, Nanjing 210096, China and the Purple Mountain Laboratories, Nanjing 211111, China (e-mail: yong\_zeng@seu.edu.cn).}}

\maketitle

\begin{abstract}
Channel knowledge map (CKM) is a novel technique for achieving environment awareness, and thereby improving the communication and sensing performance for wireless systems. A fundamental problem associated with CKM is how to construct a complete CKM that provides channel knowledge for a large number of locations based solely on sparse data measurements. This problem bears similarities to the super-resolution (SR) problem in image processing. In this letter, we propose an effective deep learning-based CKM construction method that leverages the image SR network known as SRResNet. Unlike most existing studies, our approach does not require any additional input beyond the sparsely measured data. In addition to the conventional path loss map construction, our approach can also be applied to construct channel angle maps (CAMs), thanks to the use of a new dataset called CKMImageNet. The numerical results demonstrate that our method outperforms interpolation-based methods such as nearest neighbour and bicubic interpolation, as well as the SRGAN method in CKM construction. Furthermore, only 1/16 of the locations need to be measured in order to achieve a root mean square error (RMSE) of 1.1 dB in path loss.
\end{abstract}


\section{Introduction}
\IEEEPARstart{F}{or} the sixth-generation (6G) mobile communication networks, the expansion of frequency bands, together with the adoption of extremely large-scale multiple-input multiple-output (MIMO), renders traditional methods for real-time channel state information (CSI) acquisition more costly and time-consuming, prompting the pursuit of innovative approaches for CSI acquisition \cite{10430216}. Channel knowledge map (CKM) \cite{9373011} is a promising technique to address such challenges. CKM provides location-specific channel knowledge that is crucial for enhancing environment-awareness, and hence may significantly improve the communication and sensing performance. For example, the authors in \cite{10287775} proposed an environment-aware hybrid beamforming technique based on CKM, which drastically reduces the real-time training overhead. By leveraging user location information, this method significantly improves the effective communication rate, even in the presence of moderate location errors. A training-free beamforming scheme was proposed in \cite{10108969}, which designs optimal active and passive beams based on the location and environmental information provided by CKM. Besides, CKM-enabled environment-aware networks can realize communication-aware trajectory planning to avoid blind spots for network-connected ground or aerial robots.

A fundamental challenge for CKM-enabled environment-aware communication and sensing is developing effective methods for CKM construction. In \cite{10530520}, the authors employed an analytical model to construct channel gain map (CGM), where channel modeling parameters were estimated from measured data, thereby allowing the full CGM to be generated through the model. In addition to model-based methods, the authors of \cite{7817747} proposed a data-driven approach for constructing CKM, employing Kriging interpolation to build the shadowing map from a limited set of observations. Meanwhile, the rise of deep learning-based CKM construction methods has sparked a growing demand for specialized CKM datasets. The dataset of path loss and time of arrival (ToA) radio maps \cite{yapar2022dataset} encompasses simulated path loss/received signal strength (RSS) and ToA radio maps. The  CKMImageNet dataset \cite{Wu2024CKMImageNet} provides location-tagged numerical channel data alongside visual imagery, offering a comprehensive view of both the channel and environment. This integration not only supports the validation of various communication and sensing algorithms but also enables CKM construction using advanced computer vision (CV) techniques.

The emergence of these specialized datasets has advanced researchers' investigations into the aforementioned data-driven methods. RadioUNet \cite{9354041} leverages a physical simulation dataset to produce path loss estimations. In \cite{10130091}, the authors devised a sophisticated network architecture grounded in conditional generative adversarial networks (cGANs), engineered to synthesize detailed radio maps. The authors of \cite{jin2024i2i} treat channel knowledge as a 2-D image, framing the CKM estimation as an image-to-image (I2I) inpainting task that predicts channel knowledge at a specific location. Although extensive research has explored data-driven CKM construction with neural networks, these approaches require physical maps or other supplementary information as input.

Super-resolution (SR) aims to convert a low-resolution (LR) image into a corresponding high-resolution (HR) image with improved visual quality \cite{10.1145/3390462}. Compared to LR images, HR images exhibit a higher pixel density and more detailed textures, resulting in greater reliability. Nowadays, deep learning has become the dominant method in the field of super-resolution. SRCNN \cite{7115171}, as one of the earliest models to apply deep learning techniques to super-resolution, achieved peak signal-to-noise ratio (PSNR) that significantly surpassed traditional methods. In addition to convolutional neural networks (CNNs), researchers have also explored the potential of GANs \cite{9151093} and residual networks (ResNets) \cite{li2018multi}. All of these approaches have achieved impressive results in image super-resolution. 

\begin{figure}[htbp]
\centering
\includegraphics[width=3in]{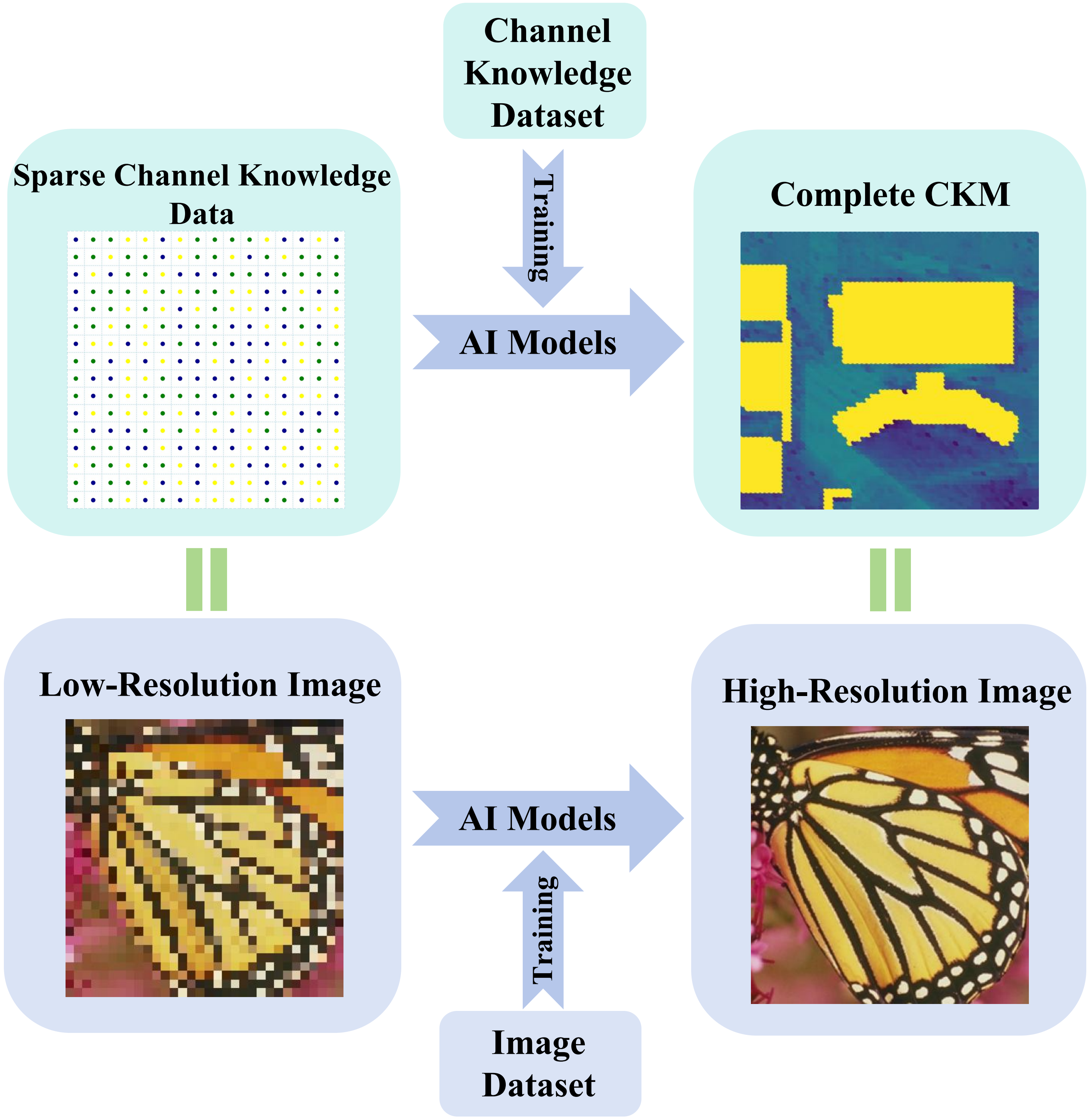}
\caption{Analogy between CKM construction using sparse data and image super-resolution.}
\label{F:1}
\end{figure}

In this letter, we draw an analogy between image super-resolution and CKM construction, as illustrated in Fig.~\ref{F:1}. This enables us to address the CKM construction problem from an image processing perspective, leveraging advanced SR algorithms from the CV domain to obtain the complete CKM from sparse data. The most straightforward image SR algorithms are interpolation methods, which directly perform linear interpolation based on existing pixel values, such as nearest neighbour (NN) and bicubic interpolation. However, these methods often result in significant errors. Other advanced image SR networks, such as GANs, emphasize the visual quality of images, which may not be suitable for channel knowledge inference. Considering the unique features and metrics of CKM, we employ SRResNet to obtain the complete CKM based on sparse channel measurements. Unlike image processing, which highlights texture detail and perceptual realism \cite{ledig2017photo}, our approach prioritizes the accurate prediction of channel knowledge. Therefore, we employ mean square error (MSE) as the loss function. By utilizing the novel CKMImageNet dataset, our method can construct not only traditional path loss maps but also channel angle maps (CAMs). Numerical results demonstrate the effectiveness of our proposed strategies across various CKM quality metrics. 

\section{System Model}\label{sec:SystemModel}
Consider a base station (BS) aiming to construct a CKM based on the sparse measurements within its coverage area. Depending on the application requirements, various types of channel knowledge may be included in the CKM, such as path loss, AoA and angle of departure (AoD), etc. Without loss of generality, we assume that the desired CKM contains $w\times h$ locations each carrying $c$-dimensional channel knowledge.  

In practical applications, dense and frequent channel measurements are costly. We assume that only sparse measurements are collected at $w'\times h'$ uniformly distributed locations, which is much smaller than $w\times h$. We define the expected HR CKM as a vector $\textbf{x}^{*}\in\mathbb{R}^{wh\times c}$ and the measured sparse data as $\textbf{y}^{*}\in\mathbb{R}^{w^{\prime} h^{\prime} \times c}$. Therefore, $\textbf{y}^{*}$ can be considered a uniform sampling of $\textbf{x}^{*}$. We use a diagonal matrix $\textbf{H} \in \mathbb{R}^{w^{\prime}h^{\prime} \times {wh}}$ with diagonal elements that are either 0 or 1 to denote the sampling process, i.e.,

\begin{equation}
	\textbf{y}^{*}=\textbf{H}\textbf{x}^{*}.
\end{equation}

Since $w^{\prime}h^{\prime} \ll wh$, obtaining $\textbf{x}^{*}$ from $\textbf{y}^{*}$ constitutes an underdetermined problem, making it impossible to find a unique solution without additional information. Fortunately, by utilizing the CKM dataset, we can employ a neural network trained through supervised learning to reconstruct $\textbf{x}^{*}$ from $\textbf{y}^{*}$. 

\begin{figure}[htbp]
\centering
\includegraphics[width=3in]{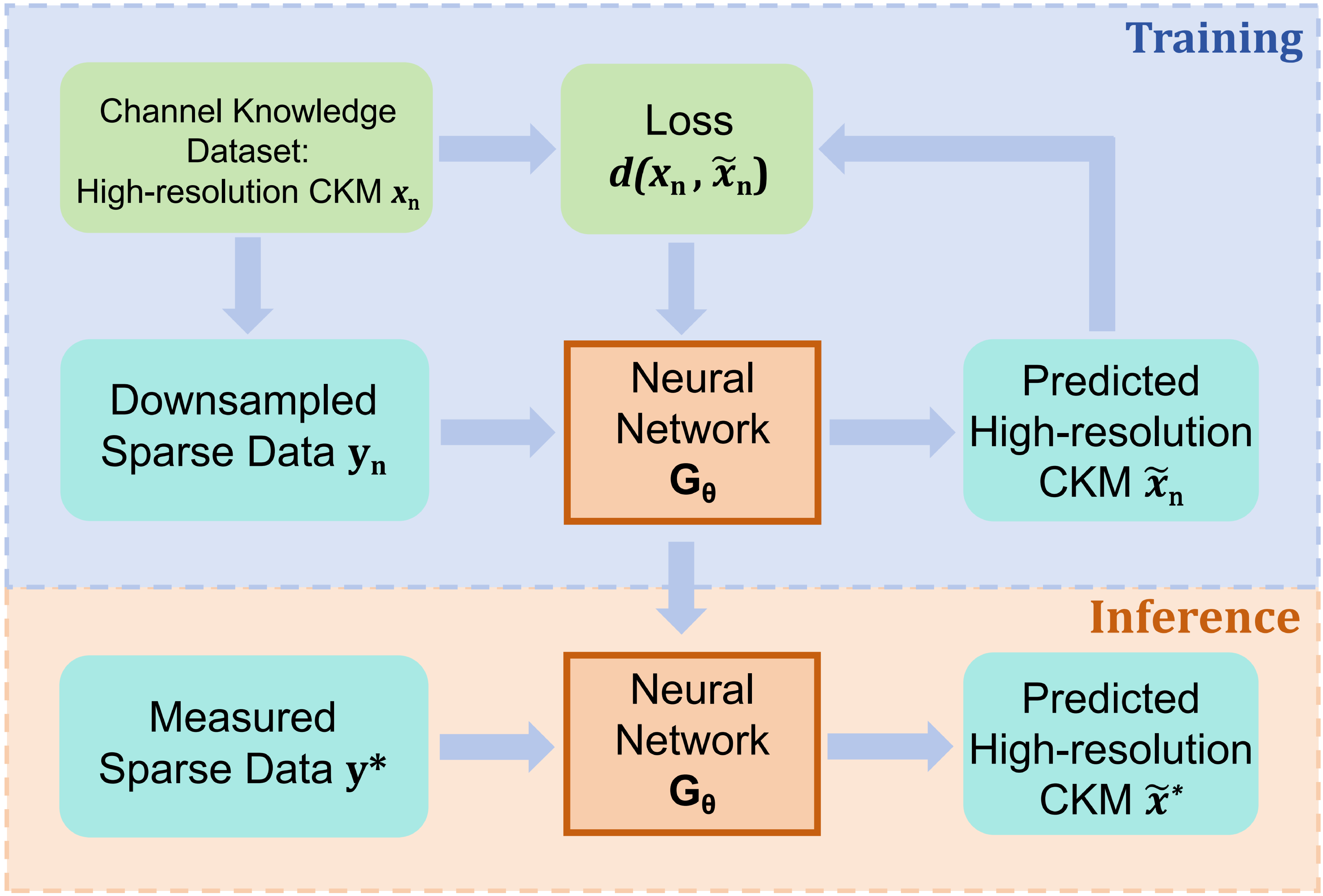}
\caption{Illustrating of the training and inference phases of deep learning-based CKM construction.}
\label{F:system}
\end{figure}

Fig.~\ref{F:system} illustrates the training and inference processes of the neural network $\mathit{G_\theta}$ for CKM construction. During the training process, we sample sparse data $\textbf{y}_n$ from the high-resolution CKM $\textbf{x}_n$ provided by the CKM dataset. The error between the predicted CKM $\widetilde{\textbf{x}}_n$ and the ground truth $\textbf{x}_n$ is calculated to update the network parameters. During inference, the measured sparse data $\textbf{y}^{*}$ is directly input into the trained network to construct the HR CKM.

The error between the constructed CKM and the ground truth can be defined as

\begin{equation}\label{3}
	\mathcal{E} = \mathit{d}(\widetilde{\textbf{x}}_n , \textbf{x}_n),
\end{equation}

where $\mathit{d(\cdot,\cdot)}$ denotes the loss function. We utilize channel knowledge-specific metrics such as MSE at the pixel level and root mean square error (RMSE) at the channel data level to assess the error in channel knowledge. Additionally, we consider other common metrics used in the field of image processing to evaluate image quality, such as PSNR, structural similarity index measurement (SSIM) \cite{1284395}, and the emerging learned perceptual image patch similarity (LPIPS) \cite{8578166}.

\section{Architecture and Methodology}\label{sec:Architecture}
In this section, we discuss the details of the network architecture and the methodology applied during the training process. 

\subsection{Network Architecture}
We leverage the SRResNet \cite{ledig2017photo} architecture, which contains 1,549,462 parameters, as our model for performing image super-resolution. Although the SRGAN, proposed concurrently, excels in generating realistic SR images, we opted for SRResNet due to its superior performance in pixel-level restoration, which aligns more closely with our goal.

The core of SRResNet consists of a sequence of five residual blocks, each meticulously designed with a uniform architecture. These blocks feature two $3\times3$ convolutional layers with 64 feature channels, complemented by two Batch Normalization (BN) layers, and utilize ParametricReLU (PReLU) functions for activation. With LR images as input, the network executes a series of convolutional operations and ultimately integrates two sub-pixel convolutional layers  to upscale the image resolution. This process results in the generation of SR images with resolutions that are significantly higher than those of the input LR images.

\subsection{Training Process}
Our overarching objective is to train the SRResNet $\mathit{G_\theta(\cdot)}$, parametrized by $\theta$, capable of constructing the corresponding HR counterpart $\textbf{x}_n$ for given LR input $\textbf{y}_n$, where $\theta=\{W_{1:L};b_{1:L}\}$ denotes the weights and biases of the $L$-layer SRResNet. Based on our training objective, we select MSE as the loss function to optimize the parameters. Thus, for CKM images $\textbf{x}_n$ of size $w\times h$, $n=1,\dots,D$, our objective is to solve

\begin{equation}\label{7}
	\hat{\theta} = \arg\min_\theta\sum_{n=1}^{D}\frac{1}{wh}(\|G_\theta(\textbf{y}_n)-\textbf{x}_n\|)^2,
\end{equation}

where both $\textbf{x}_n$ and the downscaling counterpart images $\textbf{y}_n$ of size $w^{\prime} \times h^{\prime}$, with $n=1,\dots,D$, serve as the network input.

The training procedure of SRResNet is shown in Algorithm~\ref{algo:SRResNet}. We made several adjustments to the training procedure of the original SRResNet. The batch size is set to 32, the number of iterations is 100,000, the learning rate is 0.001. Additionally, the image cropping operation has been removed to maximize the utility of the sparse measured data.

\begin{algorithm}
	\caption{The training procedure for CKM construction.}
	\label{algo:SRResNet}
	\begin{algorithmic}
		\STATE \textbf{Input:} \raggedright HR CKM $\textbf{x}_n$, upscaling-factor $k$, learning rate $\gamma$,
		\STATE \hspace{0.95cm} batchsize $m$, total number of iterations $T$, 
		\STATE \hspace{0.95cm} number of images in the training set $D$.
		\STATE
		\STATE Calculate the number of iterations per epoch $N = D / m$ and the epoch number $E = T / N$; 
		\STATE Randomly initialize parameters $\theta^{(0)} = \{W^{(0)}_{1:L}; b^{(0)}_{1:L}\}$;
		\FOR{$t = 1, 2, \dots, E$}
		\FOR{$n = 1, 2, \dots, N$}
		\STATE Randomly select $\textbf{x}_n$ of batchsize $m$;
		\STATE Sample $\textbf{y}_n$ from $\textbf{x}_n$ using the factor of $k$;
		\STATE Input $\textbf{y}_n$ into SRResNet;
		\STATE Calculate the loss $d(G_{\theta^{(n-1)}}(\textbf{y}_{n-1}), \textbf{x}_{n-1})$;
		\STATE Calculate gradients $\nabla_{\theta^{(n-1)}} d(G_{\theta^{(n-1)}}(\textbf{y}_{n-1}), \textbf{x}_{n-1})$;
		\STATE \raggedright Update parameters $\theta^{(n)} = \theta^{(n-1)} - \gamma \nabla_{\theta^{(n-1)}} d(G_{\theta^{(n-1)}}(\textbf{y}_{n-1}), \textbf{x}_{n-1})$
		\ENDFOR
		\ENDFOR
		\RETURN Trained parameters $\theta^{(T)} = \{W^{(T)}_{1:L}; b^{(T)}_{1:L}\}$.
	\end{algorithmic}
\end{algorithm}

The number of pixels in image $\textbf{x}_n$ corresponds to the number of information-carrying locations in each CKM. Through $k\times$ SR, the number of locations in constructed CKMs is $k^2$ times that in sparse measurements. Thus, we have the following equation

\begin{equation}\label{8}
	\mathit{k}=\sqrt{\frac{w h}{w' h'}}.
\end{equation} 

In this letter, we set $k=2, 4, 8$ and $16$, which indicates that the sparse data required to construct a HR CKM is only $1/4, 1/16, 1/64$ and $1/256$ of the total data.

\section{Numerical Results}\label{sec:Results}
In this section, we describe the dataset used and present the numerical results.

\subsection{Simulation Setup}
\subsubsection{Datasets}
We first utilize the RadioMapSeer\footnote{https://radiomapseer.github.io/} dataset, from which we choose 21,000 path loss maps simulated through the dominant path model (DPM) for training and testing. The training dataset consists of 20,000 randomly selected path loss maps, while the test dataset comprises the remaining 1,000 maps. We ensure that the BS locations in the test dataset are not included in the training dataset, allowing us to evaluate the model's generalization capability. 

We also train and test our model on the CKMImageNet\footnote{https://github.com/Darwen9/CKMImagenet} dataset. CKMImageNet provides both location-tagged numerical channel data and visual images, offering a holistic view of the channel and environment. Built using commercial ray tracing software, CKMImageNet captures electromagnetic wave propagation in various scenarios, revealing the relationships among location, environment and channel knowledge \cite{Wu2024CKMImageNet}. The dataset encompasses a wide variety of channel parameters, including path loss, delay, AoA and AoD, etc. The vast majority of existing datasets focus solely on path loss, while the CKMImageNet dataset incorporates angle information such as AoA and AoD. We randomly select 11,064 path loss maps from CKMImageNet for the training set and 1,000 distinct path loss maps for the test set. Compared to path loss, angular information in the channel often exhibits a more complex distribution, making it more challenging to construct. Therefore, we employ AoA maps in CKMImageNet to evaluate whether SR algorithms can also be utilized to construct CAMs. The AoA map training set consists of 11,866 images, while the test set includes an additional 1,000 distinct images. All the images in the CKMImageNet dataset are of size $128\times128$. 

\subsubsection{Pixel-data Correlation}
\begin{figure}[H]
\centering
\subfloat[]{\includegraphics[width=1.5in]{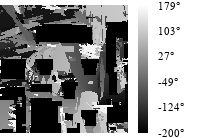}}
\hfil
\subfloat[]{\includegraphics[width=1.5in]{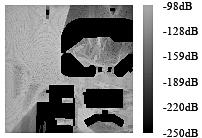}}
\caption{Mapping between pixel values in CKM images and channel knowledge. (a) AoA map. (b) Path loss map.}
\label{F:CKM_colorbar}
\end{figure}

Before delving into the numerical results, it is essential to elucidate the relationship between the image information and the channel knowledge contained within the CKM. Fig.~\ref{F:CKM_colorbar} illustrates how the grayscale pixel values in the CKM images are mapped to the corresponding channel knowledge.

In the CKMImageNet dataset, the simulated path loss ranges from -250 dB to -50 dB and is linearly mapped to the grayscale range of [0, 255]. For the AoA maps, angles ranging from -180° to 180° are linearly mapped to the same grayscale range, while the angle for the building locations is set to -200°. In the RadioMapSeer dataset, the linearly mapped range for path loss is from -147 dB to -47 dB.

\subsection{Performance and Stability}
\subsubsection{Accuracy}
Under the condition of 4$\times$ SR, we compared the performance of SRResNet with the NN interpolation, bicubic interpolation and the SRGAN. The PSNR, SSIM, and LPIPS values are shown in Table \ref{T1} and Table \ref{T2}, which demonstrate the excellent performance of the proposed SRResNet-based CKM image SR method. Moreover, the significant reduction in MSE and RMSE further indicates the great potential of SRResNet in CKM construction. Based on the numerical results from the simulation on the RadioMapSeer dataset, only $1/16$ of the locations need to be measured in order to achieve a RMSE of $1.1$ dB in path loss.

\begin{table}[h]
\caption{Performance on RadioMapSeer dataset(4$\times$ SR).}\label{T1}
\centering
\begin{tabular}{c c c c c c}
			~ & PSNR & SSIM & LPIPS & MSE(pixel) & RMSE(dB) \\ \hline
			nearest & 25.46 & 0.8557 & 0.1307 & 195.8 & 6.283 \\ 
			bicubic & 27.20 & 0.8627 & 0.2719 & 131.5 & 5.182 \\  
			SRGAN & 39.82 & 0.9856 & \textbf{0.0071} & 7.690 & 1.409 \\  
			SRResNet & \textbf{41.53} & \textbf{0.9900} & \textbf{0.0071} & \textbf{5.460} & \textbf{1.063} \\ 
\end{tabular}
\end{table}

\begin{table}[h]
\caption{Performance on CKMImageNet dataset(4$\times$ SR).}\label{T2}
\centering
\begin{tabular}{ c c c c c c}
			~ & PSNR & SSIM & LPIPS & MSE(pixel) & RMSE(dB) \\ \hline
			nearest & 20.17 & 0.6827 & 0.1618 & 842.0 & 24.97 \\ 
			bicubic & 21.72 & 0.6983 & 0.3159 & 579.4 & 20.89 \\ 
			SRGAN & 25.42 & 0.8272 & 0.0460 & 214.8 &  13.12 \\ 
			SRResNet & \textbf{30.87} & \textbf{0.8747} & \textbf{0.0317} & \textbf{76.64} & \textbf{7.684}\\ 
\end{tabular}
\end{table}

\begin{figure}[htbp]
\centering
\includegraphics[width=3in]{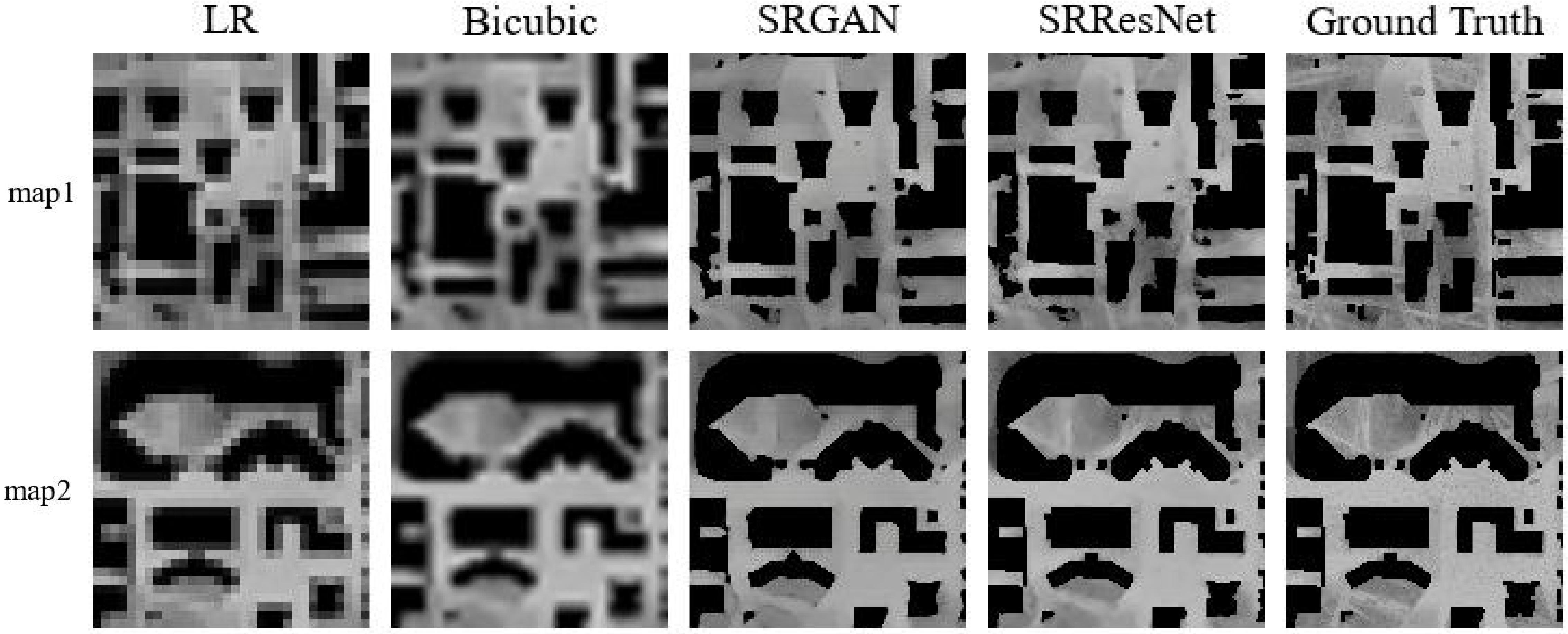}
\caption{Visualization of 4$\times$ super-resolved path loss maps. (CKMImageNet)}
\label{F:visualization1}
\end{figure}

In addition to the numerical results, we also provide visualizations to intuitively demonstrate the application of SR algorithms in CKM construction. Fig.~\ref{F:visualization1} presents the visualized SR results of selected maps from the test dataset of CKMImageNet. To provide a more intuitive demonstration of the SR effect, in this letter, we replace the LR images with the results of NN interpolation. It can be discerned that both SRGAN and SRResNet exhibit the ability to produce images that are visually almost indistinguishable from the ground truth, demonstrating a high degree of fidelity in their outputs.

\subsubsection{Performance under Different SR Factors}

\begin{figure}[htbp]
\centering
\includegraphics[width=3in]{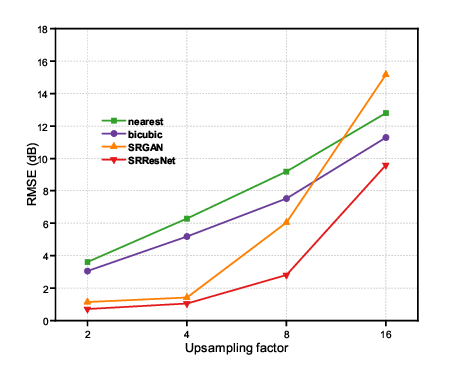}
\caption{Comparison of RMSE under different SR factors. (RadioMapSeer) }
\label{F:data}
\end{figure}

\begin{figure}[htbp]
\centering
\includegraphics[width=3in]{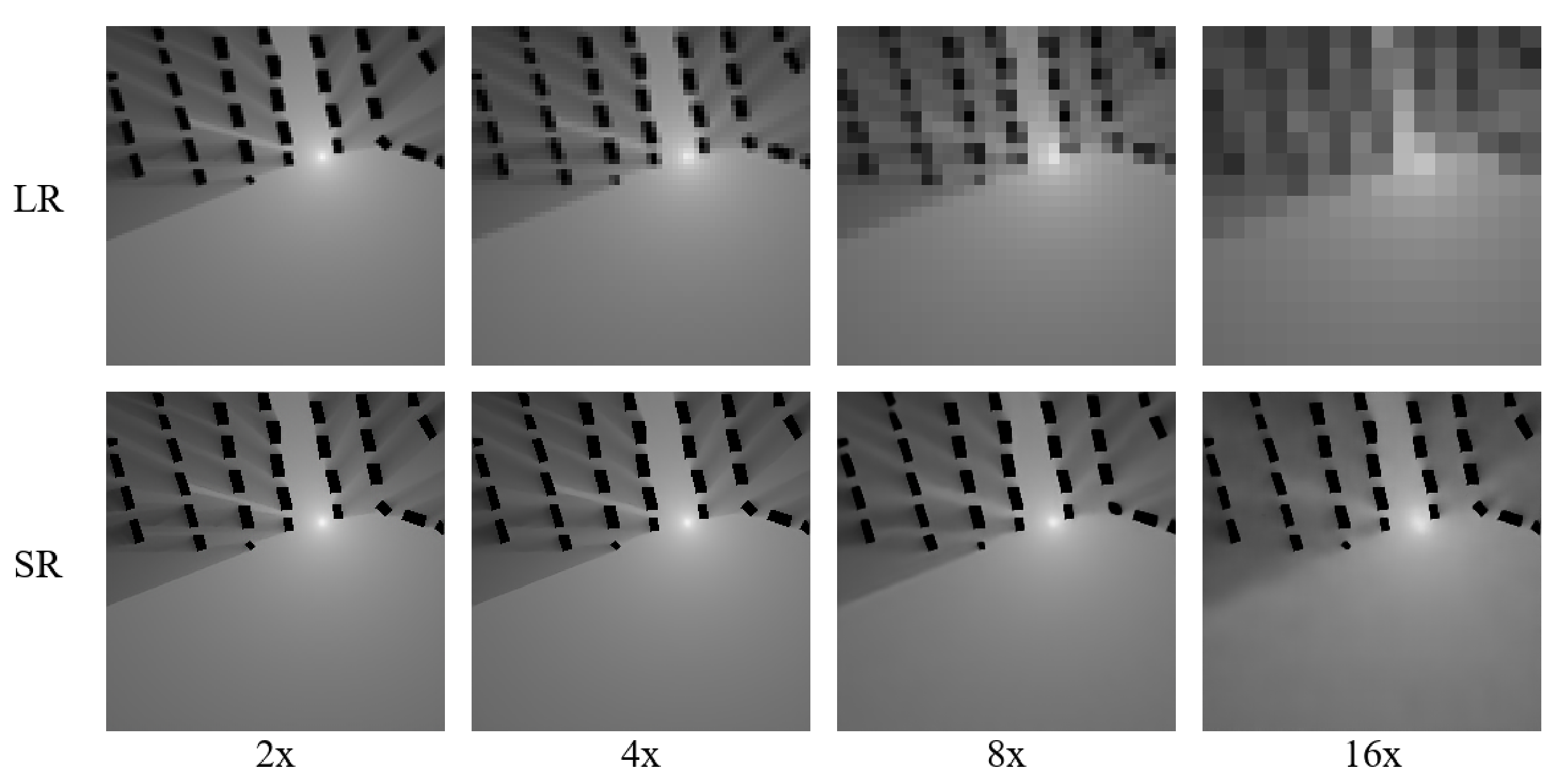}
\caption{Visualization of LR path loss maps (NN interpolated) and constructed maps under different SR factors.}
\label{F:visualization2}
\end{figure}

In image processing applications, 2$\times$, 4$\times$, and 8$\times$ are common SR factors \cite{10.1145/3390462}, but higher factors may be required for channel data prediction. In Fig.~\ref{F:data}, we compare the RMSE performance of different algorithms at various SR factors and find that SRResNet consistently outperforms other methods across all factors. For example, a RMSE of approximately $9.6$ dB in path loss was achieved with $1/256$ expected data. Furthermore, SRResNet exhibits more stable performance compared to SRGAN. We also demonstrate the visualization results under different SR factors in Fig.~\ref{F:visualization2}.

\subsubsection{Performance of AoA prediction}

\begin{figure}[htbp]
\centering
\includegraphics[width=3in]{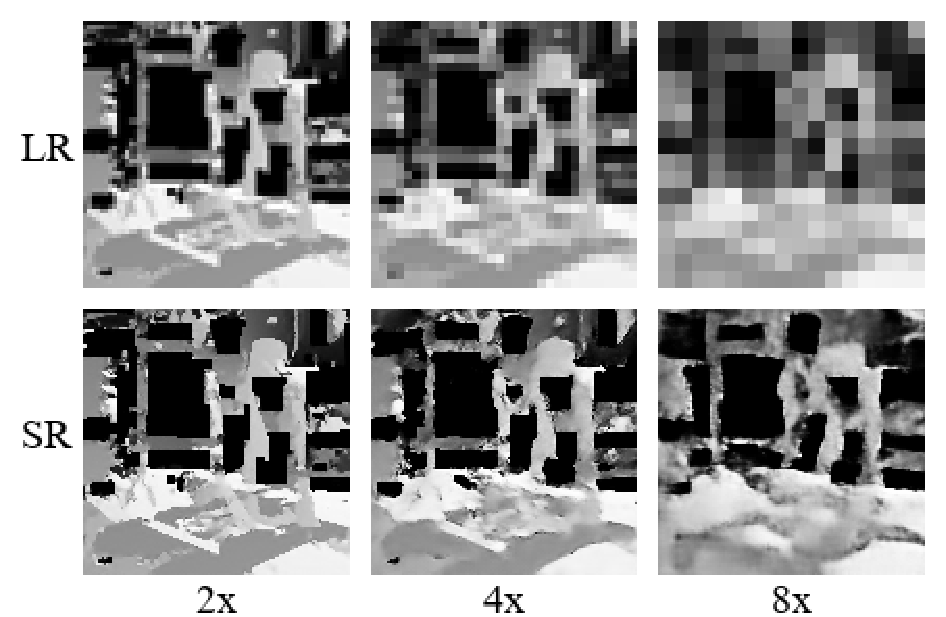}
\caption{Visualization of LR AoA maps (NN interpolated) and constructed maps under different SR factors.}
\label{F:aoa}
\end{figure}

Fig.~\ref{F:aoa} illustrates the performance of SRResNet method on constructing AoA maps, demonstrating that SR techniques are also effective for CAM construction. Comparing the visual results in Fig.~\ref{F:visualization2} and Fig.~\ref{F:aoa}, we observe that under the same SR factor, SRResNet performs better on path loss maps than on AoA maps. This discrepancy arises because AoA maps, with their inherently more complex and nuanced features, present a greater challenge for construction using CNN compared to path loss maps, when using the identical training scheme.

\subsection{Discussion}
Through simulations on two different datasets, we demonstrate that our method requires only sparse observation data as input to construct CKM, and it outperforms traditional linear interpolation algorithms in both visual quality metrics and channel knowledge accuracy. Moreover, the SRResNet method is capable of constructing not only HR path loss maps but also AoA maps. Conducting cross-dataset transfer learning and constructing more complex channel knowledge represent significant challenges and will be important areas for future work.

\section{Conclusion}\label{sec:Conclusion}
In this letter, we propose a novel CKM construction approach based on an image SR algorithm, namely SRResNet, which demonstrates superior performance in accurately and efficiently constructing complete CKMs directly from sparse data. Through simulations, we demonstrate the excellent performance of SRResNet method compared to other methods across different factors and CKM datasets. Furthermore, our approach can also be applied to the construction of HR CAMs, and the generalization to other types of channel knowledge is straightforward. In the future, we will consider the CKM construction with noisy and non-uniformly distributed channel measurements. 

\bibliographystyle{IEEEtran}
\bibliography{IEEEabrv,bibliography}

\end{document}